\documentstyle[graphics,graphicx,times]{mn2e}
\input{psfig}

\def\parn{\par\noindent}

\def\los{line of sight}
\def\loss{lines of sight}

\def\dv{\Delta\, v}
\def\pa{{\parallel}}
\def\pe{{\perp}}
\def\Om{\Omega_{\rm m}}

\def\Ol{\Omega_{\Lambda}}
\def\hMpc{$h^{-1}$Mpc}
\def\kms{km~s$^{-1}$}

\def\R#1{{\mathrm{#1}}}
\def\Sec#1{{Section~\ref{s:#1}}}
\def\Eq#1{{Eq.~\ref{e:#1}}}

\def\EQN#1{\label{e:#1}}        
\def\Fig#1{{Fig.~\ref{f:#1}}}

\def\d#1{{\R{d}{#1}}}
\def\mdot{\!\cdot\!}
\def\HI{H~{\sc i}}

\begin{document}

\title[Matter clustering from the Lyman-$\alpha$ forest]{
The correlation of the Lyman-$\alpha$ forest in close 
pairs and groups of high-redshift quasars: clustering 
of matter on scales 1-5~Mpc\thanks{Based on observations carried out at the 
European Southern Observatory  with UVES (ESO programme No. 65.O-299) and 
FORS2 (ESO programme No. 66.A-0183) on the 8.2~m VLT-Kueyen 
telescope operated at Paranal Observatory; Chile.}
}

\def\inst#1{{${}^{#1}$}}

\author[Rollinde et al.]{ E.~Rollinde\inst{1}, P.~Petitjean \inst{1,2}, 
C.~Pichon\inst{1,3,4}, S.~Colombi\inst{1,4}, B.~Aracil\inst{1},
\newauthor V.~D'Odorico\inst{1}, M.G.~Haehnelt\inst{5}  \\
$^1$ Institut d'Astrophysique de Paris, 98 bis boulevard
        d'Arago, 75014 Paris, France \\
 $^2$ LERMA, Observatoire de Paris, 61, avenue de l'observatoire F-75014
        Paris, France \\
        $^3$ Observatoire de Strasbourg, 11 rue de l'Universit\'e,
 67000 Strasbourg, France.\\
 $^4$ Numerical Investigations in Cosmology (N.I.C.), CNRS, France\\
 $^5$ Institute of Astronomy, Madingley Road, Cambridge CB3 0HA, England
}
\date{Typeset \today ; Received / Accepted}
\maketitle

\begin{abstract}
We study the clustering of matter in the  intergalactic medium from the
Lyman-$\alpha$ forests seen in the spectra of pairs or groups of 
$z$~$\sim$~2 quasars observed with FORS2 and UVES at the VLT-UT2 Kueyen ESO 
telescope. The sample consists of five pairs with separations 
0.6, 1, 2.1, 2.6 and 4.4~arcmin and a group of four quasars with 
separations from 2 up to 10~arcmin. This unprecedented data set allows 
us to measure the transverse flux correlation function 
for a range of angular scales. Correlations are clearly detectable  
at separations  smaller than  3~arcmin. The shape and correlation
length of the transverse correlation  function on these scales is in good 
agreement with those expected from absorption by the photoionized warm 
intergalactic medium associated with the filamentary and sheet-like
 structures predicted in CDM-like models for structure formation. 
At larger separation no significant correlation is detected. 
Assuming that the absorbing structures are randomly 
orientated with respect to  the line of sight, the comparison of 
transverse and longitudinal correlation lengths  constrains the cosmological
parameters (as a modified version of  the Alcock \& Paczy\'nski test).
The present sample is too small to get significant constraints.
Using $N$-body simulations, we investigate the possibility to constrain 
 $\Ol$ from future larger samples of QSO pairs  with similar
separations.
The observation of a sample of 30 pairs at 2, 4.5 and 7.5 arcmin 
should constrain the value of $\Ol$ at $\pm$ 15 \%\ (2$\sigma$ level).
We  further use the observed spectra of the group of four  quasars, to search for  
underdense regions in the intergalactic medium. We find  a quasi-spherical
structure of reduced absorption with radius 12.5 $h^{-1}$Mpc which 
we identify as an underdense region. 
\end{abstract}


\section{Introduction}

The  intergalactic medium (IGM) is  revealed  by  the numerous  H~{\sc  i}
absorption  lines seen  in the  spectra  of distant  quasars, the  so-called
Lyman-$\alpha$ forest. For a long time these absorption lines  have
been  believed  to be  the signature  of  discrete and compact intergalactic
clouds photoionized by the UV-background (e.g., Sargent et al. 1980). However,  
$N$-body simulations (Cen et  al. 1994; Petitjean, M\"ucket \& Kates 1995; 
  Zhang, Anninos \& Norman  1995; Hernquist  et  al.   1996;   M\"ucket  al  al.  1996;
Miralda-Escud\'e et al. 1996; Bond  \& Wadsley 1998; Theuns et
al. 1998) and analytical works (e.g., Bi \& Davidsen 1997; Hui \& Gnedin 1997)
together with the  first determination of the approximate size 
of the absorbing structures from observation of QSO pairs 
(Bechtold et al. 1994; Dinshaw et al. 1994) established 
a new paradigm. The Lyman-$\alpha$ forest is now generally believed to arise 
instead from spatially extended density fluctuations of moderate amplitude
 in the continuous intergalactic medium. The baryons  thereby
follow the dark matter distribution on scales larger than the Jeans length.
Observations of the  Lyman-$\alpha$ forest
can thus be used  to constrain  structure formation  models and cosmological  
parameters.   Croft et
al. (2000)  e.g. used hydrodynamic simulations to investigate  the
relation between  the flux power spectrum  
along the \los\  
and the three-dimensional linear dark matter power spectrum  $P(k)$. The 
slope and amplitude of $P(k)$ derived from observed absorption 
spectra is similar to  those of  popular CDM models.  
%
%

\begin{figure*}
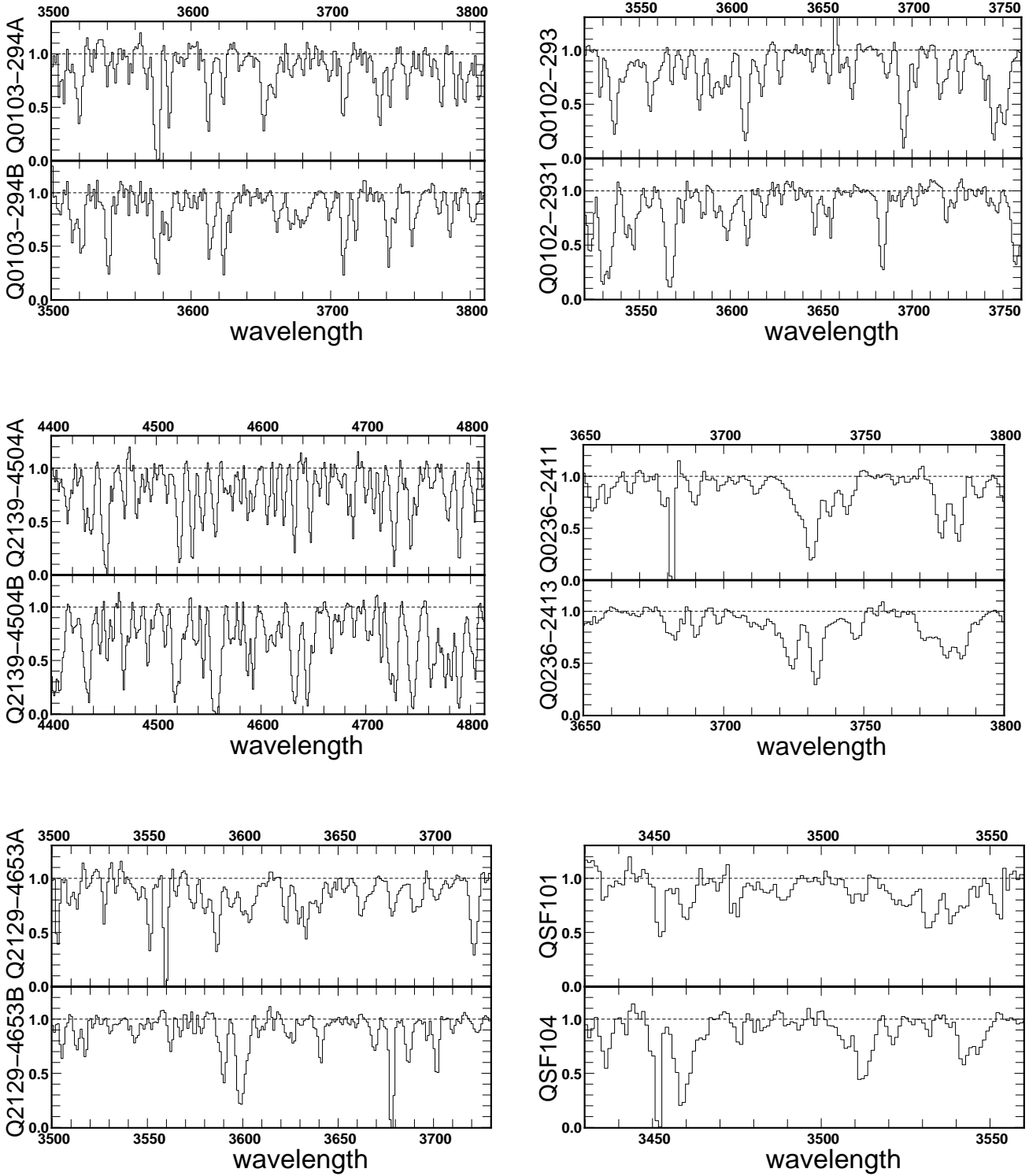

\unitlength=1cm
\begin{picture}(15,20)(1,0)
\put(0,14){\psfig{width=8cm,angle=-90,figure=MC461f1a.ps}}
\put(9,14){\psfig{width=8cm,angle=-90,figure=MC461f1b.ps}}
\put(0,7){\psfig{width=8cm,angle=-90,figure=MC461f1c.ps}}
\put(9,7){\psfig{width=8cm,angle=-90,figure=MC461f1d.ps}}
\put(0,0){\psfig{width=8cm,angle=-90,figure=MC461f1e.ps}}
\put(9,0){\psfig{width=8cm,angle=-90,figure=MC461f1f.ps}}
\end{picture}
\caption{Spectra of the observed pairs of QSOs (see Table~1). The separations
are (from left to right and upper to lower panels): 2.1, 9.5, 0.6, 2.6, 2.1, 4.4 arcmin.}
\label{f:pairs}
\end{figure*}

%
\parn A more direct method to  constrain the three-dimensional 
distribution of the  intergalactic medium is the use of 
absorption spectra along multiple \loss.  This enables the study of
the  correlation between the Lyman-$\alpha$ forests observed along 
adjacent lines of sight  to quasars
with small projected separations in the sky. Such  observations
contain valuable  
information in both the longitudinal and  transverse directions.

\par\noindent
The observed Lyman-$\alpha$ forest is most of the time decomposed
in discrete absorption features (see however e.g. Pichon et al. 2001).
Along the line of sight, absorptions with log~$N$(H~{\sc i})~$>$~14
are strongly correlated for velocity separations
$\Delta v$~=50-100~km~s$^{-1}$ ($\simeq$ 0.5-1 $h^{-1}$~Mpc
com., $h=H_0/100$; all distances are in comoving
units in this paper ; 
Cristiani et al. 1997; Khare et al. 1997; Kim et al. 1997).
In the transverse direction, observations of pairs of quasars are 
rare. 
From the statistics of coincident absorption features along the
lines of sight to multiple images of lensed QSOs 
(Mc Gill 1990; Smette et al.  
1992, 1995) and/or close pairs or triplets of QSOs (Bechtold et al. 1994; 
Dinshaw et al. 1994; Fang et al. 1996;
Charlton et al. 1997; Crotts \& Fang 1998; D'Odorico et al. 1998; 
Petitjean et al. 1998; Young, Impey \& Foltz 2001; 
Aracil et al. 2002), a  typical size  of  300-400 $h^{-1}$  kpc  has been  
inferred assuming spherical absorbers. However, 
the scatter in the measurements is large due to rather poor 
statistics. These large  sizes  reflect
more likely the correlation length of smaller 
absorbing structures rather than the real size 
of continuous  independent objects.   
Indeed, correlation of discrete clouds has been claimed along
adjacent lines of sight of separations up to 0.7 $h^{-1}$~Mpc at $z>$2 
(Crotts 1989; Dinshaw  et al.  1995; Crotts \& Fang 1998) and even 
2 $h^{-1}$~Mpc  (Liske et al. 2000) and 30  $h^{-1}$~Mpc (Williger et 
al. 2000).  Finally, a strong correlation signal is seen up
to separations  of about 2  $h^{-1}$~Mpc at  $z$~$\sim$~1 by Young  et al.
(2001) and Aracil et al. (2002).  \par\noindent
We present here the analysis of intermediate and high spectral resolution 
high-quality observations of  
five QSO pairs and a group of four quasars. They span separations from 0.6 to 9.5 
arcmin ($\simeq$ 0.2 to 3.3 $h^{-1}$~Mpc proper, 
using  $\Om=0.3$ and
$\Ol=0.7$) at $z\simeq  2.2$ (see Table 1). Details of the observations are  presented
in Section~2. In  Section~3 we define and   measure the flux 
correlation function along the \los\  and in the transverse  
direction. In Section~4 we discuss  how the flux correlation function  
is influenced by redshift distortions. We then discuss further 
how  to  use the Alcock 
\& Paczy\'nski (1979) test to constrain the cosmological parameters 
characterizing the geometry of the Universe (essentially $\Lambda$ 
for a flat Universe)  by comparing  the relationship between  the 
line of sight correlation function and the transverse correlation
function  which  is cosmology dependent. In  the last section  we use 
observations of  a quasar quadruplet to  search for underdense regions 
(defined as regions of reduced absorption) in the 
intergalactic medium.
%
%

\section{Observations and Simulation}
\label{s:data}
\begin{figure}
\centerline{\includegraphics[width=7.5cm]{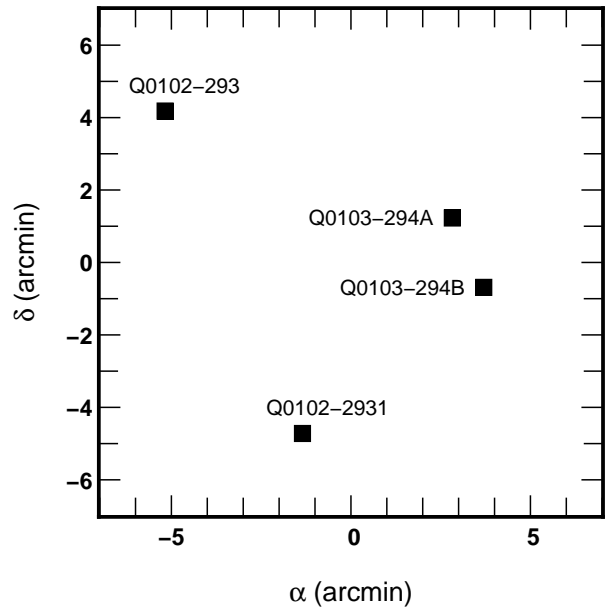}}
\caption{The group of four quasars at redshift $z\sim2$
located  in a  field of  10~arcmin aside. It corresponds 
to the first two pairs in Table~{\ 1}.  This specific 
configuration allows us to study directly  the  three-dimensional 
structure of the intergalactic medium and to search for 
underdense regions in \Sec{quadruplet}.}
\label{f:quadruplet}
\end{figure}

 \begin{table}
 \begin{tabular}{|c c|c|c|}
 \hline
 \hline
 quasar & $z_{\rm em}$ & wavelength range & separation\\
        &          & \AA              & arcmin\\
 \hline
 Q0103-294a & 2.15    & 3500-3810 & 2.1\\
 Q0103-294b & 2.15   &  3500-3810 &     \\
 \hline
 Q0102-2931 & 2.22   &  3520-3760 & 9.5  \\ 
 Q0102-293 & 2.44   &   3485-4000 &       \\
 \hline
 \hline
 Q2139-4504a & 3.05  & 4155-4813 &0.6 \\ 
2139-4504b & 3.26  & 4400-5030 & \\
 \hline 
Q0236-2411 & 2.28 & 3550-3860 & 2.6 \\
 Q0236-2413 & 2.21 & 3650-3800 & \\ 
\hline 
 Q2129-4653a & 2.18 &  3500-3730 & 2.1 \\ 
Q2129-4653b  & 2.16 & 3500-3765 &       \\ 
 \hline
 QSF101 & 2.22 &  3315-3805 & 4.4 \\ 
QSF104  & 2.02 & 3430-3560 &      \\
 \hline
 UM680 & 2.144&  3200-3800& 1.0 \\ 
UM681 & 2.122&           &       \\
 \hline 
 \end{tabular} 
\caption[]{
Some characteristics of  the observed pairs.
 The normalized    spectra   are   presented    in  
 \Fig{pairs}.   The wavelength  range, from  the  Lyman-$\beta$ emission
 (if above 3500  \AA) to 3000 km~s$^{-1}$ from the quasar emission redshift,
is  indicated.  The first four
 quasars  belong  to the  same  field  as  shown in  \Fig{quadruplet}.  This
 quadruplet is equivalent  to six pairs, which yields a  total of 11 pairs.}
\label{t:paire}  
 \end{table}

The  quasars have typical magnitudes in the range  
$m_{\rm  V}$~=~18.5-20.5  and
redshifts  $z_{\rm em}$~$>$~2. They were selected  to  belong to 
pairs  or groups  of  quasars with separations of a 
few arcmin. The spectra were obtained  with FORS2 mounted on VLT-UT2
Kueyen ESO telescope using the grism GR630B  and a 0.7~arcsec slit.   
The spectra were reduced  using the standard
procedures available in  the context LONG of the  ESO data reduction package
MIDAS.  Master bias and flat-fields were produced from day-time 
calibration.  Bias subtraction  and flat-field  division  were 
performed  on science  and calibration images. A correction  for 
2D  distortion was applied.  The sky spectrum  was evaluated
in  two windows on both  sides  of the  object  along the  slit
direction   and  subtracted  on  the  fly   during  the  optimal
extraction of the object.  The spectra  were  then wavelength  
calibrated  over the  range 3400~$<$~$\lambda$~$<$~5000~\AA.
The  final   pixel  size   is  1.18~\AA~
and the resolution 
is   $R$~=~1400 or  FWHM~=~220~km~s$^{-1}$ at
3800~\AA\ (or 2.3 pixels).  The exposure times have  been adjusted to obtain 
a typical signal-to-noise ratio of $\sim$15 at 3500~\AA.  
The sharp decrease of the detector sensitivity below
4000~\AA~ prevents  scientific analysis below 3500~\AA. 
At $\lambda$~$\sim$~4500~\AA~ the S/N ratio is usually 
larger than 70.  
The pair UM680/UM681 (56~arcsec separation) was observed with 
the high-resolution spectrograph UVES. Details of the analysis of 
the spectra of this pair are presented in another paper
(D'Odorico, Petitjean \& Cristiani 2002).
Redshifts and separations are summarized in Table~\ref{t:paire}.
Spectra are presented in \Fig{pairs}.
\par\noindent
We have used numerical simulations to estimate errors in
the correlation function (Section~\ref{s:correlation})
and test our implementation of the Alcock
\& Paczy\'nski test (Section~\ref{s:APtest}). 
A large LCDM $N$-body simulation was run with the Particle~-~Mesh
(PM) code described in Pichon et al. (2001). The simulated box
is  a cube of 100 \hMpc\ size, and involves 16 million particles
on a 1024$^3$ grid to compute the forces. 
This large box size  is required to probe the linear regime
at $z\sim 2$ (Section~\ref{s:APtest}). The wavelength range of 
the Lyman-$\alpha$ region in the observed spectra ($\sim 250$ \AA) 
corresponds to approximately twice the length of our 
simulation box (200 \hMpc). The initial grid of the simulation 
is projected onto a 256$^3$ grid, using 
adaptative smoothing as explained in Pichon et al.:
each dark matter particle is projected on the grid with a gaussian 
profile $\rho(r)\propto \exp{(-r^2/2l^2)}$, truncated at $r=3l$,
where the typical radius $l$ is given by the mean square distance 
between the particle and its 32 nearest neighbours. 
The final spatial pixel size, 0.4 \hMpc, corresponds to approximatively 
0.47 \AA~ over the considered wavelength range and is thus smaller
than the FORS pixel size.
Cosmological parameters are $\Ol=0.7$, $\Om=0.3$, $h=0.65$
and we have chosen $\sigma_8=0.93$ consistent 
with constraints from the evolution of galaxy clusters 
(Eke, Cole \& Frenck 1996). Initial conditions are set up using the 
CDM power spectrum $P(k)$ with $n=1$ and the transfer function 
given in Efstathiou, Bond \& White (1992),
(\Eq{CDM}). 
The \HI\ optical depth along lines of sight through the
simulation box is derived as in Rollinde, Petitjean \& Pichon (2001). 
The spectra are then degraded in resolution 
and noise is added to mimic FORS observations.
%
\section{Flux correlations along and transverse to the line of sight}
 \label{s:correlation}

In this Section we discuss the flux correlations in the absorption 
spectra of our sample of quasar pairs and groups. We study 
how the correlation strength varies as the velocity distance 
along the line of sight or the angular separation between  
two lines of sight increases.
\par\noindent
As discussed above the  Lyman-$\alpha$ forest is believed to originate 
from the moderate amplitude density fluctuations in a homogeneously 
distributed intergalactic medium. The observed flux in absorption spectra 
is then a signature of the  smoothly varying intergalactic medium
density. It is thus reasonable to use the flux distribution directly 
for a statistical analysis without an identification of discrete absorption 
lines (see Petry et al. (2001) for a comparison with the line-fitting 
methodology).  We define the  unnormalized flux correlation function as:
\begin{equation} \xi_{f}(\theta,\Delta v)=\left\langle
({\cal     F}(\theta,\lambda+\Delta\lambda)-\langle{\cal    F}\rangle)({\cal
F}(0,\lambda)-\langle{\cal                  F}\rangle)\right\rangle_{\lambda}
\,,\EQN{correlation} 
\end{equation} 
where ${\cal  F}$ is the  normalized flux along  two \loss\ with
separation $\theta$  and    mean     redshift      $z$,      and
$\Delta\lambda=\lambda_0(1+z)\times\Delta   v/c$, $\lambda_0=1215.67$ \AA.
The velocity  distance corresponding to the angular separation $\theta$ 
can be written as $\Delta v_{\perp}=c\,f\,\theta$,  where  $c$ denotes 
the speed of light while $f(z)$ is a scaling factor which depends 
on the angular diameter distance. The dependence 
of $f(z)$ on cosmological parameters is discussed in \Sec{APtest}.  

\begin{figure}
\unitlength=1cm
\begin{picture}(7,8)
\put(0.4,0){\psfig{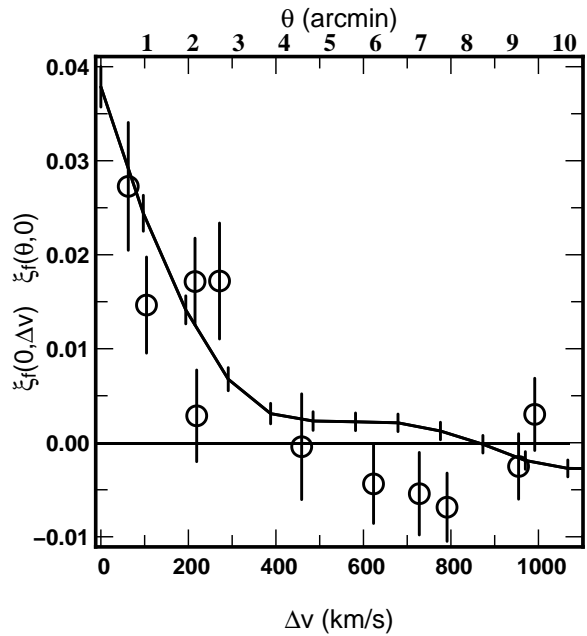}}
\put(1.38,2.45){\line(1,0){6.3}}
\end{picture}
\caption{Longitudinal (solid curve) and transverse (open circles) 
correlation function, $\xi_{f}$ of  the Lyman-$\alpha$ forest {\em vs} 
velocity (lower axis) and angular (upper axis) separation.  The definition of
$\xi_{f}$ is  given by Eq.~\ref{e:correlation}.  
Error bars (1$\sigma$) are taken from  {\em rms} of 
Monte Carlo realizations in the LCDM simulation (see text for details).
$\Om$~=~0.3  and $\Ol$~=~0.7 (1~arcmin~$\sim$~100  km~s$^{-1}$)
were assumed to relate angular separations to  velocities.}
\label{f:result}
\end{figure}

\parn The observation of each  quasar pair yields one data point of the 
transverse correlation function  ($ \Delta v=0$) at the transverse
separation between the two quasars, $\theta_{\rm i}$,
$\chi(\theta_{\rm i}) \equiv \xi_{f}(\theta_{\rm i},0)\,.$
The open circles in \Fig{result} show the transverse correlation function 
of our observed sample of {11} pairs. Note that the value of 
the transverse correlation function measured with one pair of 
quasars is averaged over the redshift range common to both  lines of  sight.  
The transverse correlation function rises strongly at angular separations 
smaller than 2-3 arcmin. For $\Om=0.3$ and
$\Ol=0.7$,  1~arcmin  corresponds to  1 $h^{-1}$~Mpc or about twice  
the Jeans length. At larger angular scales the   measurements for  the  six
pairs  with separations  between   4  and  10   arcmin  are  
consistent  with no transverse flux correlation, or possibly 
a weak anticorrelation.
\parn The 1$\sigma$ errors of the correlation function are estimated 
using  Monte Carlo realizations of 
absorption  spectra obtained from the  numerical simulation
described in the previous section. 
 The simulated absorption spectra were 
calculated along pairs/groups of lines of sight
 of constant separation randomly 
located in the YZ plane of 
the simulation. We produced several hundred realizations with   
the same separations as our sample of five pairs and one quadruplet
and calculated the transverse and longitudinal correlation 
functions for all simulated spectra.  
From this we have determined the full error matrix 
including the cross-correlation coefficients to determine 
the {\em rms} error due to cosmic variance and to look for possible 
correlations in the errors.
 We find that the error matrix is nearly diagonal for the transverse measures. 
Only the cross-correlation between the two pairs related to the quadruplet, 
Q0103-294A/Q0102-293 and Q0103-294B/Q0102-293, 
as well as the cross-correlation between Q0103-294A(B)/Q0102-2931
(see Fig.~2) are non negligible.
\parn

\begin{figure}
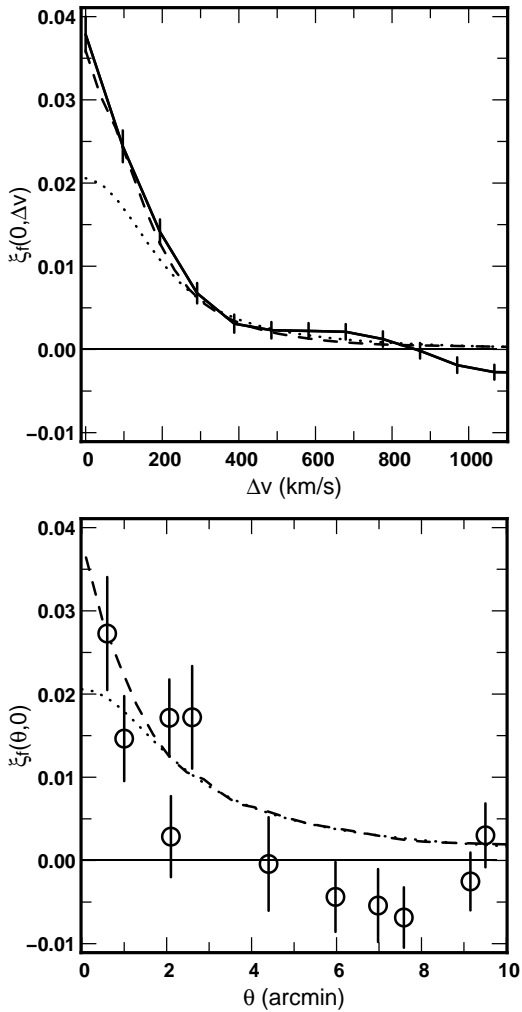

\unitlength=1cm
\begin{picture}(8,13.5)
\put(0.5,6.8){\psfig{height=6.5cm,figure=MC461f4a.ps}}
\put(0.5,0){\psfig{height=6.5cm,figure=MC461f4b.ps}}
\put(1.38,1.93){\line(1,0){5.5}}
\put(1.38,8.73){\line(1,0){5.5}}
\end{picture}
\caption{Comparison of observations, LCDM simulation and 
linear theory
for the longitudinal (upper panel) and transverse (lower panel) 
correlation functions.
Observations are plotted as in \Fig{result} with solid line
and open circles. The simulated correlation function,
plotted as dashed lines, are computed on spectra
with the same resolution and noise as FORS 
spectra. Linear theory predictions (dotted lines) correspond 
to the best fit to the LCDM simulation with $\Om=0.3$ over
the range $300<\Delta v<1000$ \kms\,; $\theta>2$ arcmin. } 
\label{f:sim_obs} 
\end{figure}

\parn The three data points at 2 arcmin seem to show a somewhat larger 
dispersion than expected from the error bars. We performed 
a Kolmogorov-Smirnov test and found that the assumption 
that these points are drawn from  a normal distribution
with {\em rms} as given by the error bar
cannot be rejected with a confidence level higher than 10 percent.
 Nevertheless, this may indicate that we have somewhat
underestimated the errors.
This point will be investigated  in the future using hydro-simulations.
 The data points at larger separation 
seem to indicate a small anti-correlation. As noticed above the 
measurements of pairs which are part of the  quadruplet  
at 5.9, 6.9,  7.5 and 9.1 arcmin are correlated. Because of 
this,
this is unlikely to be significant and the sample is probably 
just too small to get a detection of the expected weak correlation 
at large scales. If the flux were indeed anti-correlated at 
scales of a few arcmin this would not be consistent with 
a opacity distribution that traces the DM distribution 
in a simple manner (see also Meiksin \& Bouchet 1995).  
Clearly a larger sample is needed to clarify these points. 

\parn The average flux correlation along the line of sight of the twelve
spectra taken with the FORS spectrograph 
$\xi_{{f},\pa}(\Delta v)\equiv\xi_{f}(0,\Delta v)$,
is shown as the solid curve in \Fig{result}. 
Error bars are again taken as the diagonal elements 
of the error matrix obtained from the simulated spectra.   
For velocity  separations larger than our spectral resolution 
($\Delta v$~$\ge$~220~km~s$^{-1}$  at 3500 \AA) our measurement 
of the line of sight correlation function is very similar to those 
obtained from high resolution spectra  (e.g Croft et al. 2000). 
As expected, at smaller velocity separations the 
correlation strength is reduced.
\parn
In \Fig{sim_obs}, the observed correlation functions are compared to 
those obtained from the simulated spectra (dashed line). 
The agreement is good and within the error bars
(see also  Viel et al. 2002). 
\parn 
The agreement between the observed correlation functions along and 
transverse to the 
line of sight is remarkable. It rules out the possibility 
that the  line-of-sight flux correlation is caused by  peculiar velocity 
effects only and corroborates the proposition 
that the flux correlation reflects the clustering of the 
underlying matter distribution in real space. It thus further supports  
the basic paradigm that the Lyman-$\alpha$ forest is caused by  the 
fluctuating Gunn-Peterson absorption due to sheet-like and filamentary  
structures predicted by hierarchical structure formation scenarios.   
\parn 
%
\section{Applying  the Alcock \& Paczy\'nski test
to spectra of QSO pairs}

\label{s:APtest}

\parn
The detection of the transverse correlation at scales smaller than
three arcmin in our still rather small data set 
and the good agreement between the transverse and the line of sight
correlation funtions is  very encouraging for a detailed quantitative 
comparison between the transverse and the line of sight
correlation funtions. If the absorbing structures are randomly orientated 
with respect to the line of sight (which should be a trivial assumption),
such a comparison should constrain cosmological parameters because of
the cosmology dependence of the relation between  physical length 
scales measured as velocities along the line of sight and angular
distances transverse
to the line of sight (Alcock \& Paczy\'nski 1979; 
see also Hui, Stebbins \& Burles 1999; Mc Donald \& Miralda-Escud\'e 1999).
\parn
This test is often called the ``Alcock \& Paczy\'nski test''. 
It makes use of the fact that the relation 
between physical length scales measured along the line of sight as 
velocities and transverse to the line of sight as angular separations 
depends  on cosmological parameters. Any orientation independent 
measurable length scale can thus -- at least in principle -- 
constrain cosmological  parameters. In our particular case of the 
comparison of the transverse  and line of sight flux correlation 
functions, the correlation length
should be such an orientation independent length scale. Obviously if the 
shape of the correlation function is sufficiently well determined 
the whole function can be used instead. However the flux
correlation function along the line of sight will be affected by 
peculiar velocities 
which in turn depend on cosmological parameters. 
\parn Unfortunately at small scales where the correlation signal is strong
the evolution of the matter density is in the non-linear regime and
modelling becomes difficult while at large scale where the
evolution of the density field and the corresponding peculiar
velocities are well approximated by linear theory
the correlation signal is weak. As discussed earlier
we have not yet detected a correlation signal at scales larger than 
3 armin in the sample presented here. 
\parn In the following, we will first explore how the scale factor 
$f$ relating velocities to angular distances depends on cosmology and
then describe how we can model the effect of peculiar velocities using
linear theory in order to constrain $\Ol$ when samples large enough to 
measure the correlation strength at large scales become available.
 
\subsection{The scale factor $f$}

\parn 
The  relation between angular distance  and velocity 
depends on the  cosmological density parameter, $\Om$
and on the cosmological constant, $\Ol$, 
in the following way (Weinberg 1972): 
\begin{equation} \Delta v_{\perp}=c f(z) \theta\, 
\label{e:angle_dist}
\end{equation}
\parn
with 
\begin{equation} 
f(z)=\frac{H(z)/H_0}{(1+z) \sqrt{\Omega_{\rm K}}}  \Xi\left( 
\sqrt{\Omega_{\rm K}}H_0 \int_0^z \frac{\ \d z}{H(z)}\right)\,,
\EQN{equaf}
\end{equation}
\begin{equation}
H(z)=H_0\sqrt{\Om(1+z)^3+\Omega_{\rm K}(1+z)^2+\Ol}
\end{equation}
\parn 
where $\Omega_{\rm K}=1-\Om-\Ol$ and 
$\Xi$ is the  $\sinh$ function for $\Omega_{\rm K}>0$ and 
the  $\sin$ function for 
$\Omega_{\rm K}<0$. For a flat universe,
\begin{equation} f(z)=\frac{H(z) }{(1+z)}\, \int_0^z \frac{\d 
z}{H(z)}\,.
\end{equation}
\begin{figure}
\unitlength=1cm
\begin{picture}(7,7)
\centerline{\psfig{width=7cm,figure=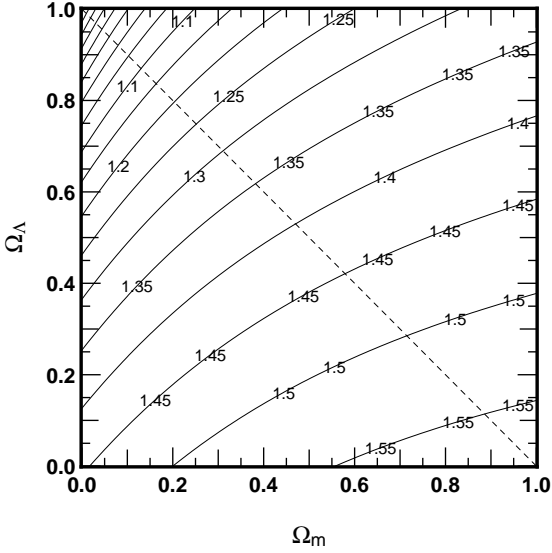}}
\end{picture}
\caption{Contours of constant $f(z)$ in the plane $\Ol - \Om$
(solid lines, separated by 0.05), at $z= 2$. $f(z)$  relates angular distance 
and velocity separation (see \Eq{angle_dist}). The dashed line 
corresponds to $\Ol\ +\ \Om=1$}
\label{f:isof}
\end{figure}

\parn   
The iso-contours of  $f$  at $z=2$ are plotted in \Fig{isof} in
the  $\Ol - \Om$ plane. They are approximately perpendicular 
to the line corresponding to a flat universe. In the following, 
we assume that $\Ol + \Om = 1$. $f(z)$ varies by about 
50 percent for $\Om$ in the range  $0.1<\Om<0.9$.

\subsection{Modelling the effect of peculiar velocities with linear theory}
\parn 
As discussed above the flux correlation function along the
line of sight will be affected by peculiar velocities. We will 
use linear theory and a CDM power spectrum to model their 
effect. The use of linear theory will introduce some systematic 
biases but as we will demonstrate using numerical simulations 
these are smaller than our observed uncertainties. 
A more detailed treatment will probably be needed 
to take advantage of  more precise measurements of the transverse 
correlation function expected from future larger samples of QSO pairs. 
\parn 
In linear theory the redshift space correlation function 
of matter is related to the
real space correlation as follows (Kaiser 1987; Matsubara \& Suto 1996)~:

\begin{eqnarray}
\xi(\dv_\pe,\dv_\pa)&=&\left(1+\frac{2}{3}\beta+\frac{1}{5}\beta^2
         \right)
         \xi_0(s)-
 \nonumber \\ &&
 \left(\frac{4}{3}\beta+
\frac{4}{7}\beta^2\right)
         \xi_2(s)P_2(\mu)+
 \nonumber \\ &&
         \left(\frac{8}{35}\beta^2\right)\xi_4(s)P_4(\mu) ~,
\label{e:linearcorrelation}
\end{eqnarray}     
where $s=\sqrt{\dv_\pa^2+\dv_\pe^2}$, $\mu=\dv_\pa/s$,
$P_{\rm l}(\mu)$ are
the Legendre polynomials, and
\begin{equation}
\xi_{\rm l}(s)=b^2 \times 4\pi \int_0^\infty \d k k^2 P(k)  \,
j_{\rm l}\left[k\, s\, (1+z)\times h/H(z)\right] ~.
\label{e:xil}
\end{equation}
The parameter $b$ allows for a linear bias between the observed 
object (the flux) and the underlying mass.
The functions $j_{\rm l}(x)$ are the spherical Bessel functions, and
$P(k)$ is the power spectrum of the mass fluctuations, $k$ is 
in units of $h$~Mpc$^{-1}$. The parameter
$\beta$ gives account of the peculiar velocity effect. It
is related to the bias parameter $b$ by $\beta \simeq 
\Om(z)^{0.6}/b$.
For the  Lyman-$\alpha$ forest and 
the linear expansion case, this bias can be written as  
\begin{equation}
b=2-0.7*(\gamma-1)
\label{e:bias}
\end{equation}
where $\gamma$ specifies the temperature-density relation
$T\propto \rho^{\gamma-1}$ (Hui 1998). 
\parn 
For the power spectrum P(k) we assume the following  parameterization 
of cold dark matter models (Efstathiou et al. 1992),
 
 \begin{equation}
P(k)=\frac{k^n}{\left(1+(\alpha_1 q+(\alpha_2 q)^{1.5}+(\alpha_3 q)^2)^{\nu}\right)^{2/\nu}}\,,
 \label{e:CDM}
\end{equation}
\parn where $q=k/\Gamma$, $\Gamma=\Om h$ and
 $\alpha_1=6.4$, $\alpha_2=3.$, $\alpha_3=1.7$,
 $\nu=1.13$.
Thermal broadening, peculiar velocity smearing and some non-linear effects lead 
to a smoothing of the flux on small scales. This can be
reasonably modelled by multiplying the power spectrum given in
\Eq{CDM}  by a factor of the form, 
\begin{equation} 
 \exp\left(-[k\,.\,(1+z)\,h/H(z)]^2 v_{\rm s}^2 /2\right)\,,
\EQN{smoothP}
\end{equation}
where $v_{\rm s}$ is a free parameter. 
\parn 
Using the above formulae\ \Eq{angle_dist} to \Eq{smoothP}, 
and an overall normalisation of the correlation function $\xi_f=C^{nt}\xi$,
one can estimate $\beta$ and $f$ and therefore $\Ol$  
by a simultaneous fit of the longitudinal and 
transverse correlation functions of the flux.

\subsection{Tests with numerical simulations}

\begin{figure}
\centerline{\includegraphics[width=7cm]{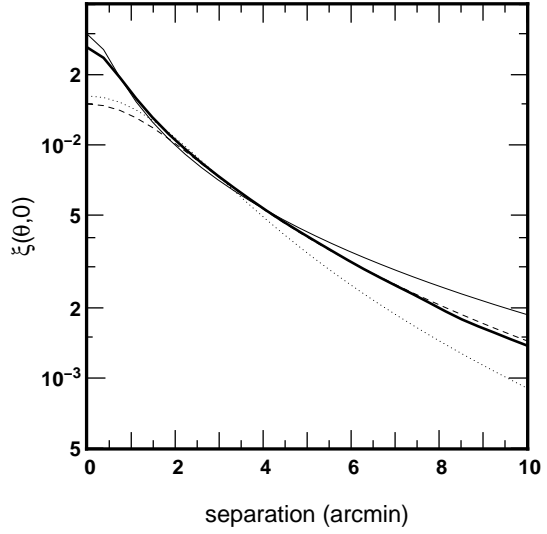}}
\caption{Application of the Alcock \& Paczy\'nski test to simulated
pairs of  quasars. The transverse correlation function, 
averaged over the simulated box (thick line; $\Ol=0.7$)  is compared to
predictions of linear theory.
The solid, dashed and dotted curves are the best fitting predictions 
of linear theory  for  $\Ol$~=~0.9, 0.7 and 0, respectively.  }
\label{f:resultat_AP_sim}
\end{figure}
\begin{figure}
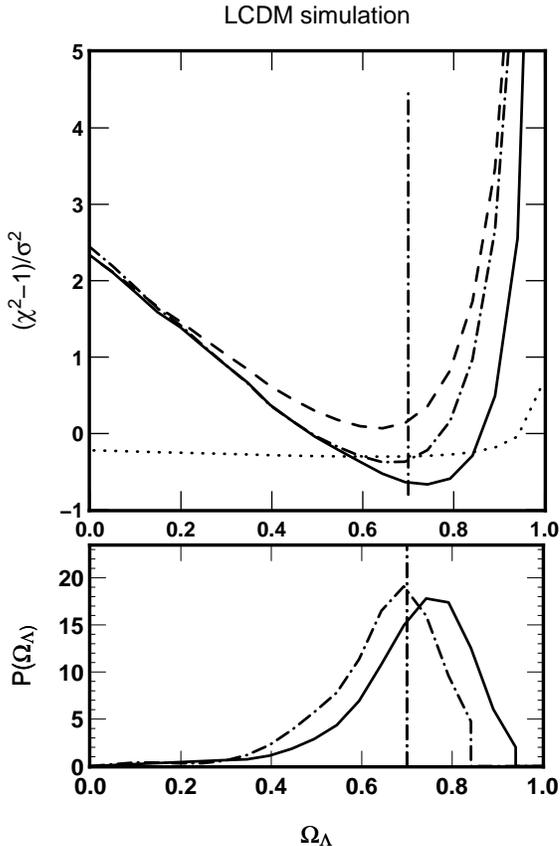

\unitlength=1cm
\begin{picture}(7,11)
\put(0,4){\psfig{width=7cm,figure=MC461f7a.ps}}
\put(0,0){\psfig{width=7cm,figure=MC461f7b.ps}}
\end{picture}
\caption{Results of the Alcock \& Paczy\'nski test applied to simulated
pairs of  quasars. 
{\em Upper panel:} 
The minimum $\chi^2$ over [$b$, $v_{\rm s}$, $C^{nt}$],
averaged over 300 realizations, 
is plotted versus $\Ol$ for three different simulated samples;
$\sigma$ is the variance of a gaussian distribution.
The dotted and dashed lines correspond to 
a sample of  1 and 30 times the number of pairs in the
current sample in the linear domain (one quadruplet and three pairs).
The solid  line corresponds to 30 pairs at 2 arcmin and 
30 pairs at 4.5 arcmin. An addition of 30 pairs at 7.5 arcmin
yields  the dot-dashed line. 
The vertical line indicates the true value of $\Ol$.
{\em Lower panel: } Distribution of the values of $\Ol$ corresponding
to the minimum $\chi^2$ for the same 300 realizations. 
The lines correspond  to the same samples as in the upper panel.}
\label{f:chi_AP_sim}
\end{figure}

The linear theory predictions for the correlation functions  are 
defined by a set of  three parameters ${\Omega_\Lambda,\ v_{\rm s},\
b}$, and the constant, $C^{nt}=\xi_f/\xi$,
for a  power spectrum index assumed to be $n=1$. 
We want to 
fit simultaneously  the  longitudinal and transverse 
correlation functions of the observed flux to the 
linear theory prediction 
to derive constraints on $\Ol$, minimizing a 
function 
 $\chi^2$, varying $v_{\rm s},\ b\ {\rm and\ } C^{nt}$.
To check this procedure, we apply it to 
mock absorption spectra obtained from our numerical simulation.

\parn 
The procedure we suggest here
 can be applied only at scales where linear theory is valid. 
Therefore we first find the domain of validity in
velocity and
angular separation 
by comparing  the linear prediction for $\Ol=0.7$ 
to results from our simulation.
The best fit of the linear theory prediction to the
correlation functions obtained from the simulated spectra
is overplotted on \Fig{sim_obs} (dotted line)
 and corresponds to
$b=1.47$, $v_{\rm s}=100$ \kms\ 
and $C^{nt}=0.0055$.
It appears that along the line of sight linear theory is a good approximation
for $v > 300$ \kms. In the transverse direction, 
this is the case for separations larger than $\sim$2 arcmin. 
As expected the scale below which  non-linear effects become  important 
is somewhat larger along the line-of-sight due to the effect of 
peculiar velocities.
\parn
In the following we restrict our analysis
to scales where linear theory is a good approximation.
We search for the value $0\le \Ol\le 1$ for which 
the linear theory prediction fits the correlation  functions 
obtained from  the simulated spectra  best varying the values of 
$b$, $v_{\rm s}$ and $C^{nt}$. 
For each set of parameters [$\Ol$, $b$, $v_{\rm s}$, $C^{nt}$],
 we compute the function 
$\chi^2~=~{\left[(\xi_{lin}-\xi_{sim})^{\bot} \mdot {\bf M}^2
 \mdot (\xi_{lin}-\xi_{sim})\right]}/N$.
$\xi_{lin} \mbox{\ are\ }\xi_{sim}$ are the longitudinal and transverse correlation
function predicted by linear theory and the simulation, respectively.
${\bf M}$ is the error matrix computed from our simulation
 (see \Sec{correlation}) 
 and $N$ is the number of pixels in the ranges considered.
As expected  $\Ol$ is mostly constrained by the 
transverse correlation function. In \Fig{resultat_AP_sim}
the linear theory predictions for  $\Ol$~=~0, 0.7  and 0.9 
are overplotted on the transverse correlation function obtained from 
the simulated spectra. Our procedure appears to recover 
the correct value of $\Ol$.
\parn
To verify this quantitatively, 
we perform 300 Monte Carlo  realizations of this test with samples 
of simulated absorption spectra containing 1 and 30  
times the number of spectra of the  subset of our observed sample 
with angular separation in the linear domain
(the quadruplet and three pairs,
see Table~\ref{t:paire}).
For each experiment,
$\chi^2$ is minimized over [$b,\, v_{\rm s},\, C^{nt}$]
for each $\Ol$.
The averaged $\chi^2$ is plotted versus $\Ol$ in the
upper panel of \Fig{chi_AP_sim}.
The dotted curve is for 
a sample of the same size as our observed sample while 
the dashed curve is for a sample 30 times this size
(the other curves are discussed in the next section).
Note that since the factor $f$ varies more strongly 
for $\Ol$ close to 1 (see \Fig{isof}), the upper limit
is more strongly constrained than the lower limit.
Our implementation of the  Alcock \& Paczy\'nski test
for the simulated spectra does indeed successfully recover 
the correct value of $\Ol$. 
We shall discuss 
below the distribution of the values of $\Ol$ which
minimize $\chi^2$.
\parn The sample of 30 times the current set of 
observations may not be efficient in terms of observationnal
strategy. We discuss in the next section the analysis of 
future observed  samples
consisting of pairs only.

\subsection{Constraints on $\Ol$ from future observations}

\parn 
The  dotted curve in Figure 7 shows that the current sample is 
not large enough to get a significant constraint on $\Ol$  even if 
we had marginally detected the weak correlation at large scales. 
The observed sample is instead consistent 
with no and possibly even a weak anti-correlation.
However, as discussed above  this is probably due to the the 
presence of a quadruplet and the resulting  correlations in the errors.
\parn
Close quasar pairs with arcmin separation are  rare
and we now discuss the question of how many pairs do 
we need to derive significative  constraint on $\Ol$.
For this, we shall consider a sample consisting 
of pairs only (without additional correlations due to the
presence of a quadruplet).
 We find first that the results obtained with
the current sample, as described in Sec~4.3, 
are  very similar  to the results with
uncorrelated  pairs with separations
 2, 2.1, 2.6, 4.4, 6 and 7.5  arcmin.
However, pairs at 2 arcmin are rare. 
We thus want to minimize the required number of pairs with small  separation. 
The solid curve  in \Fig{chi_AP_sim}
shows the $\chi^2$ distribution  
for a sample of 30 pairs with 
a separation of  2 and 30 pairs with 4.5 arcmin. 
The corresponding probability distribution is shown 
in the lower panel of \Fig{chi_AP_sim}.
The distribution has a maximum   
at $\Ol\simeq 0.76$ with a variance of 0.1 and 
a cut-off at $\Ol \simeq 0.94$. 
 Additional pairs with large 
separation, which  are available in larger number, lower
the  upper limit. Three sets of 30 pairs at 2, 4.5 and 7.5 arcmin 
respectively, shift the peak of the distribution 
to $\Ol=0.7$ and the cut-off to $\Ol \simeq 0.85$ 
while the variance is similar (dot-dashed line in \Fig{chi_AP_sim}).
A sample of  30 pairs at 2 arcmin and,
at least, 30 pairs with larger separations 
should therefore    constraint the  value of 
$\Ol$ to about 15 percent.
\parn 
%
\section{A common underdense region in the Lyman-$\alpha$ forests of a quadruplet  of quasars}
\label{s:quadruplet}
The close quadruplet  of quasars contained in our sample is well
suited  to search 
for the possible presence of large  underdense regions in the 
intergalactic medium which should result in coincident  segments
of reduced absorption
in the    Lyman-$\alpha$  forests of the different spectra. 

\parn Previous attempts  to identify underdense regions  in the Lyman-$\alpha$  forest have searched
for regions devoid of absorption lines  along a single line of sight. They
looked for portions of spectra where the observed number of absorption lines
is  significantly smaller than  the expected  number from  synthetic spectra
(Carswell  \& Rees 1987;  Crotts 1987;  Ostriker, Bajtlik \& Duncan 1988;  Cristiani et
al. 1995; Kim, Cristiani \& D'Odorico 2001).  
As the forest is usually crowded, a statistically significant detection of
an underdense region can be achieved only  for regions of dimensions larger than 30 $h^{-1}$~Mpc. 
Underdense regions with a size  larger than  40 $h^{-1}$~Mpc at  $z$~$\sim$~2 have already  
been observed (e.g. Kim et al. 2001). They are rare however and this dimension
probably corresponds to an upper limit.
\parn 
The simultaneous presence of underdense regions along several close \loss\ will greatly
enhance the statistical significance 
of a detection. This will 
allow the detection of 
 smaller underdense regions. 
\parn
Before analysing the data we have to specify our  definition 
of an underdense region.  We define an underdense region along an individual line of sight
as a segment, $\left[ \lambda_{\rm b},\ \lambda_{\rm e} \right]$, 
where the normalized flux  ${\cal F}$, is higher than  the flux averaged over
the observed wavelength range
(${\cal F}_{\rm avg}\sim~0.85$)~:

\begin{equation}
 {\cal F}(\lambda)  \left\{
        \begin{array}{ll}
  	 > {\cal F}_{\rm avg} - \sigma(\lambda) & \mbox{if $\lambda_{\rm b}<\lambda<\lambda_{\rm e}$}\\
	 < {\cal F}_{\rm avg} + \sigma(\lambda)& \mbox{if $\lambda=\lambda_{\rm b}$ or $\lambda=\lambda_{\rm e}$}
	\end{array}
           \right. \,,
\label{e:void}
\end{equation}
 
\parn where $\sigma$ corresponds to the average value of the 
noise in the underdense region.
For multiple \loss, there is a list of $N_{\rm i}$ individual underdense regions
along each individual \los:
$\left\{ \left[ \lambda_{\rm b},\ \lambda_{\rm e} \right]^{\rm j} \right\}_{1\le {\rm j} \le N_{\rm i}}$,
where i indexes the individual \loss. 
When several individual underdense regions, along different \loss, have a range of
wavelength in common, it defines
a common underdense region. Note that, by definition, individual underdense regions along 
one \los, cannot intersect. 
\par
\begin{figure}
\centerline{\includegraphics[width=7cm,angle=-90]{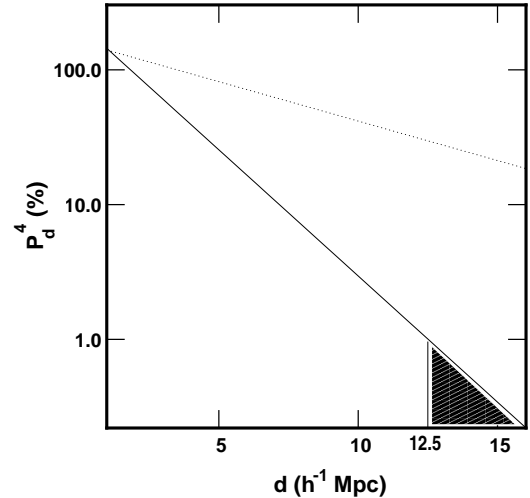}}
\caption{Probability to detect a common underdense region of size $d\ h^{-1}$~Mpc 
along four \loss\ in spectra with  random distributions of 
absorption lines.
Spectra with randomly placed lines were computed for 10000 quadruplets.
Wavelength range and S/N ratio are the same as in the observed spectra.
The corresponding  probability  for the occurence of underdense regions 
in a single \los\ is shown as dotted line.  Probabilities larger 
than 100\% correspond to more than one false detection.   
A 12.5 $h^{-1}$~Mpc underdense region was detected in the observed quadruplet; 
a  larger underdense region would occur at random with a probability  
$\simeq 2$\% (gray region).}
\label{f:proba_void}   
\end{figure}

\begin{figure}
\unitlength=1cm
\begin{picture}(7,17)
\put(0.4,11){\psfig{width=8cm,figure=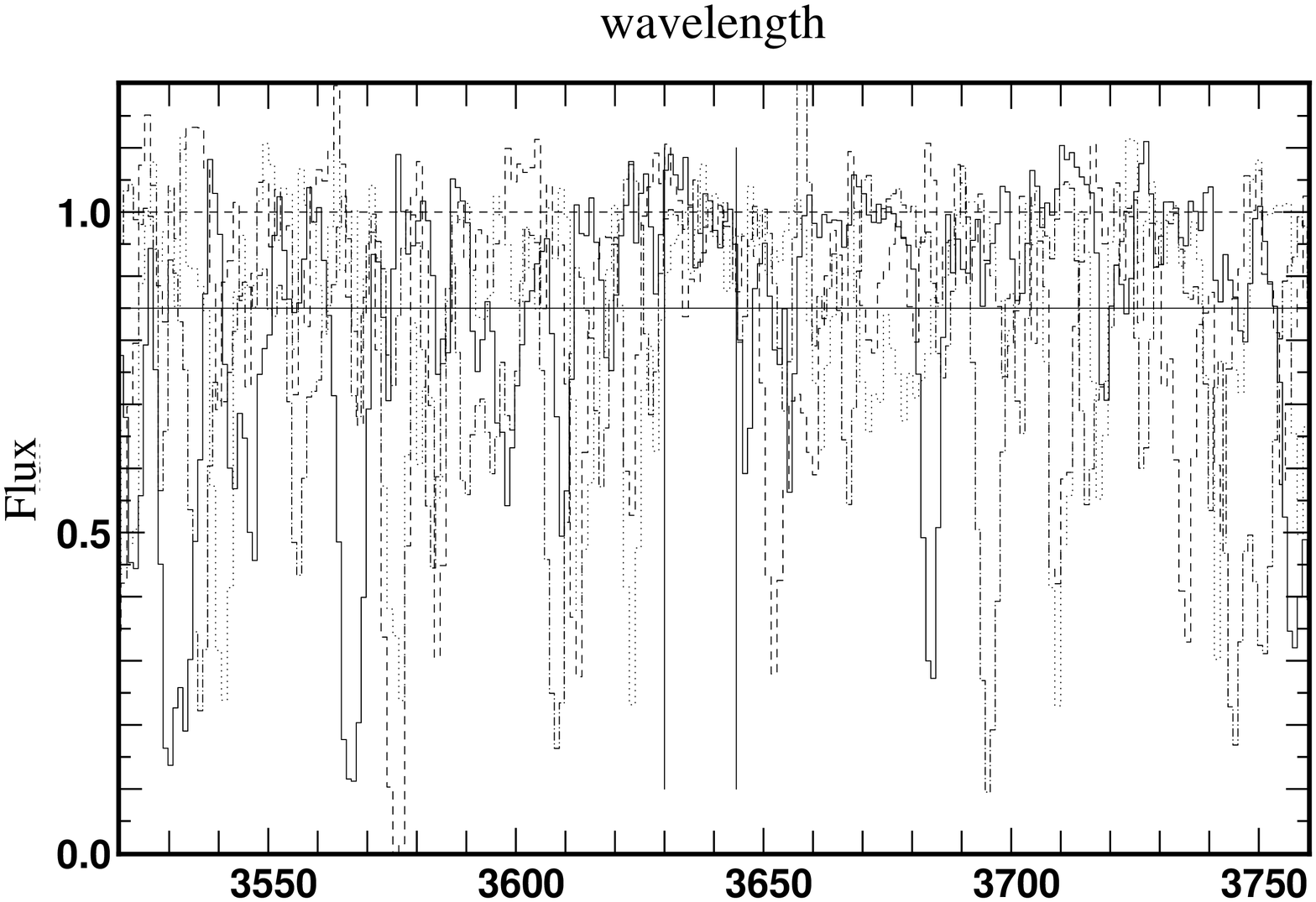}}
\put(0.5,0){\psfig{width=8cm,figure=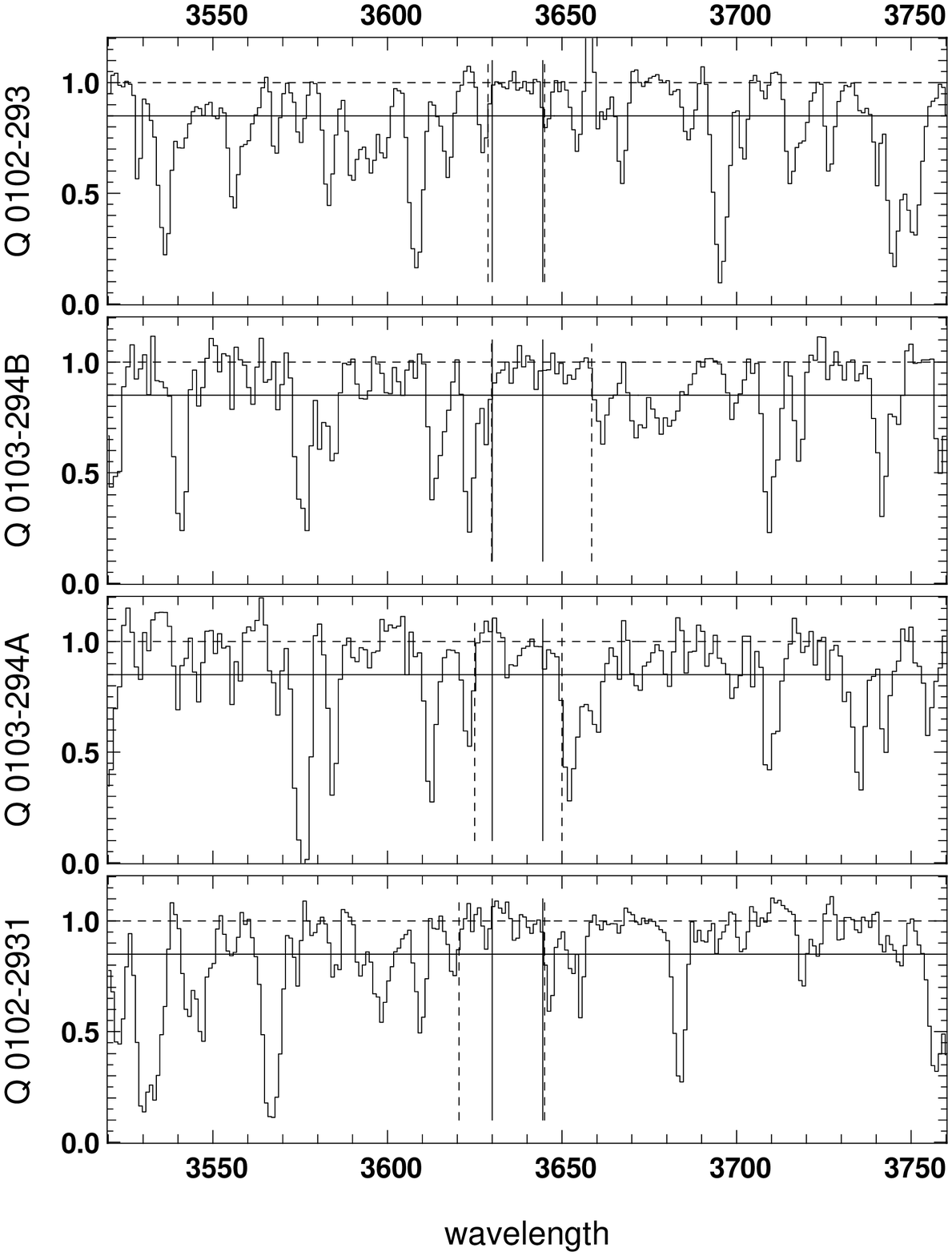}}
\end{picture}
\caption{The spectra of the close quadruplet of quasars,
\Fig{quadruplet}, centered  on a common under-dense region are plotted
in the four lower panels. Under-dense regions are defined as segments
where the flux is everywhere higher than the average flux (solid horizontal
line). The individual under-dense regions are shown with dashed
vertical lines while the common region is shown with solid vertical lines.
The same feature is 
highlighted in the upper panel 
where the four spectra are over-plotted: 
{\sl solid line} for  Q0102-2931;
{\sl dashed line} for  Q0103-294a;
{\sl dotted line} for  Q0103-294b;
{\sl dot-dashed line} for  Q0102-293.
}
\label{f:sphere}
\end{figure}

\noindent
We  then define $P_{d}^{\rm n}$  as the random
probability to have one common underdense region along n \loss\
with a size of   $d$~~$h^{-1}$~Mpc~$^($\footnote{$d_{ h^{-1}{\rm Mpc}}=c\
/ \left(100 \ \lambda_0\right)\times H_0/H\ {\rm d}\lambda_{\AA} \simeq 0.83\ {\rm d} \lambda_{\AA}$
\par $\Om=0.3$; $\Ol=0.7$; $z=2$; $\lambda_0=1215.67\, \AA$}$^)$.
The value of $P_{d}^{\rm n}$ is computed from 10000 realizations of 
n synthetic spectra along which individual underdense regions and 
common underdense regions are identified. 
The spectra are created over the observed wavelength range by randomly placing Lyman lines 
with the observed number density, column density distribution
and Doppler parameter distribution at $z\sim 2$
(e.g. Kim et al. 2002). They are degraded in resolution and noise is added 
to mimick FORS spectra.
$P_{d}^{\rm n}$  versus $d$ is plotted in \Fig{proba_void} in the case 
n=4. For comparison, $P_{d}^1$ is also indicated.
As expected, the statistical significance of a common underdense region in a quadruplet 
is 
higher than for an individual underdense region for the same size.
\parn The quadruplet consists of the pair Q 0103$-$294A/Q 0103$-$294B 
and the two quasars Q~0102$-$293 and Q~0102$-$2931. 
The separations probed are 2.1, 6, 8 and
9~arcmin. \Fig{quadruplet} shows 
the relative positions of these four quasars in 
the sky and the corresponding spectra are plotted 
in the  two upper panels of \Fig{pairs}. 
The 2-point correlation is smaller than 
the 1$\sigma$ error except for the pair Q~0103$-$294A/Q~0103$-$294B.
Each spectrum shows many individual underdense regions of different sizes. 
The most interesting feature 
is a common underdense region of\, 15~\AA\ (12.5 $h^{-1}$~Mpc).
It is located between the two vertical solid lines plotted over 
the four spectra in \Fig{sphere}, defining the wavelength range 3630$-$3645~\AA. The corresponding 
individual underdense regions are indicated with dashed lines. Their sizes are
 25, 25, 30 and 16 \AA\
($\sim$ 21, 21, 25 and 13 $h^{-1}$~Mpc) along the \los\ to Q~0102$-$2931, 
Q~0103$-$294A, Q~0103$-$294B and Q~0102$-$293, respectively. 
A broad absorption feature is present in the middle of the under-dense
segment along Q~0103$-$294A.
 One  pixel (at 3634.3 \AA),
while in agreement with \Eq{void}, is 
below the average flux. 
Note, however,  that a resolution element is actually 2.3  pixels 
and we 
consider the feature  to be above the average flux.
The common underdense region is 
clearly visible in the upper panel of \Fig{sphere} in which the four
spectra are  
plotted together.
The random probability to have a common underdense region with a size equal or larger than  
12.5~$h^{-1}$~Mpc is given by integration of $P_{d}^4$ for $d>12.5$ (shaded region 
of \Fig{proba_void}). The probability is only 
2 $\%$ which shows that the
detection of such an underdense region is 
significant.


\begin{figure}
\centerline{\includegraphics[width=5cm]{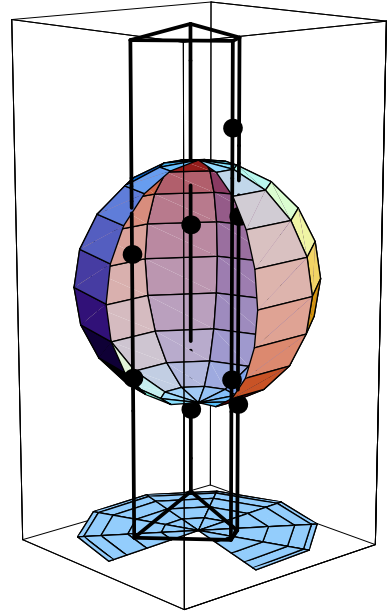}}
\caption{A schematic  three dimensional representation of the
sphere that contains a common underdense region along
a quadruplet of quasars whose spectra are shown in \Fig{sphere}.
The position of this sphere was found with a
search algorithm for three dimensional spherical 
underdense region (see text for details). The individual underdense regions
along each \los\ are marked by the black dots.
}
\label{f:sphere3d}
\end{figure}

\parn
The previous analysis did not take into account the relative positions of
the quasars in the sky. In the following, we consider underdense regions
in 3D. To detect an underdense region in Galaxy surveys (e.g. Hoyle \& Vogeley 2001),
the void finding algorithm normally identifies 
walls of galaxies with a friend-to-friend algorithm.
A void is then defined as the largest sphere that may be 
included inside one connected region.
Contrary to galaxy surveys, we do not have a complete map
of the high density regions, but individual underdense regions are fully 
defined along one \los, whereas galaxy surveys are limited by 
some magnitude threshold. 
We therefore search for the {\sl smallest} underdense region
 that includes four (in the case
of a quadruplet) individual underdense regions. This is equivalent to the requirement that 
the surface of the underdense region matches the eight  edges of the four 
segments. 
To formalize this procedure, the position of each \los\ is defined 
as $\left(x_{{\rm l}_{\rm i}},y_{{\rm l}_{\rm i}}\right)$, with 
the subscript i running from 1 to 4.
If  the 3D surface of the underdense region satisfies $V(x,y,z)=0$, then the
above requirement can be written as:
\begin{equation}
 \mbox{for ${\rm i}=1,4$}\ \left\{ \begin{array}{l}
                        V(x_{{\rm l}_{\rm i}},y_{{\rm l}_{\rm i}},{\lambda_{\rm b}}^{{\rm j}_{\rm i}}\pm {\rm d}\lambda)=0\\
       			V(x_{{\rm l}_{\rm i}},y_{{\rm l}_{\rm i}},{\lambda_{\rm e}}^{{\rm j}_{\rm i}}\pm {\rm d}\lambda)=0
 		      \end{array}
\right.\,, 
\label{e:sphere}
\end{equation}
\parn where [${\lambda_{\rm b}}^{{\rm j}_{\rm i}},\ {\lambda_{\rm
e}}^{{\rm j}_{\rm i}}$] defines a set of
four individual underdense regions as defined by \Eq{void}.
We allow for a variation  of ${\rm d}\lambda=5$\AA\ in the position of the
surface  along  each  \los.  This takes into 
account the fact that observations are in redshift space - the 
border of the underdense region may have a peculiar velocity that shifts its position
in wavelength, and that a real underdense region has probably not a simple boundary surface.
The pair  Q0102-294A/B  is  close  and highly
correlated. If the constraint given by \Eq{sphere}  is fulfilled for
either of the two \loss, it will
also be so for the other. 
 Therefore only six real constraints remain.  This set
of equations can constrain  a sphere that has four  degrees of
freedom. For simplicity we therefore search for spherical underdense regions.
Inside the sphere, the intergalactic medium
 will be under-dense along each \los\
 and on its surface it will
be over-dense  at eight  positions  identifying the  edges  
of the individual underdense regions 
(the $\lambda_{\rm b}\  \mbox{and}\ \lambda_{\rm e}$).
\parn 
When we applied this  procedure to the observed quadruplet
only one spherical underdense region was detected. It is located
at the position of the common underdense region described above. 
The radius of the sphere is also  12.5~$h^{-1}$~Mpc.
A schematic three dimensional view of this sphere and the four \loss\ 
is shown in \Fig{sphere3d}. The projection of this 3D-view in the transverse plane 
shows that the four \loss\ do not probe  the total size of the  underdense region 
in the transverse direction. 
The geometrical configuration of the quadruplet 
does not allow us  to discriminate between an ellipse 
and a sphere.

\section{Conclusion}

We have  analysed  the Lyman-$\alpha$ forest  spectra of a sample 
of close pairs  of  quasars (separations from 0.6 to 10 arcmin).  
We   concentrated on  extracting information contained 
in the measured correlation between the  Lyman-$\alpha$ forest
absorption in adjacent lines of sight. The geometrical configuration of 
a quadruplet with maximum separation  of 10~arcmin contained in the sample
has allowed us to search directly for three dimensional structures in 
the intergalactic medium. 

\parn 1.  We have detected a 12.5~$h^{-1}$~Mpc 
region
where the absorption is everywhere  below the average value, and 
 which is present in all
four   spectra of the quadruplet of quasars contained in our sample. 
The individual regions devoid of  absorption lines have a length 
of 21, 21, 25 and 13~$h^{-1}$~Mpc, respectively. 
Only 2~\% of  a sample of synthetic spectra 
with randomly distributed absorption lines have shown 
an underdense region of this size or larger by chance.  
Note that other possible interpretations include
the signature of feedback from proto-clusters (e.g. Theuns, Mo \& Schaye 2001)
or a local increase in the UV 
radiation due to the presence of a nearby source (e.g. Savaglio, Panagia \& Padovani 2002).
We also
 developed a procedure to fit a set of  close
\loss\  with spherical  under-dense regions. 
With this algorithm the four regions devoid of lines  
define a three dimensional structure of radius 12.5 $h^{-1}$~Mpc. 
\parn 2. Our sample of pairs probes a large range of angular  separations,
allowing a more accurate determination of the transverse  correlation
function than previous studies. 
Amplitude and shape of the transverse correlation function can 
be reasonably well determined up to a separation of 2-3 arcmin. 
At separations larger than  5-6 arcmin the correlation 
falls below the level which could be detected from a 
single pair of quasars. The measured transverse correlation function 
at scales smaller than 3 arcmin 
is consistent with expectations of CDM-like models of structure 
formation, where the gas is predicted to be correlated up to 
separation of a few Mpc.
At larger scales no positive correlation is detected 
and the measurements may even indicate a weak anti-correlation
which would not be consistent with the idea that the  opacity distribution 
traces the DM distribution  in a simple manner 
({\it cf} Meiksin \& Bouchet 1995). This is most likely due to the fact that 
our sample is still too small to detect the weak correlation expected
on these scales. A larger sample is needed to clarify the situation. 
We have also measured the longitudinal correlation function
and found similar results as other authors. The measured correlation
function has a well defined shape  out to a separation of  
$350-500$ km~s$^{-1}$.                         
While measured transverse separations are directly related to 
physical distances, measured line of sight separations 
are affected by peculiar velocities. Nevertheless the 
measured transverse and longitudinal correlation functions
have similar shape and correlation length 
for reasonable choices of cosmological parameters.
 The longitudinal correlation function must thus 
indeed measure spatial density correlation on scales of a few
Mpc  as  expected if the fluctuating
Gunn-Peterson effect due to the filamentary and sheet-like structures 
predicted by CDM-like cosmogonies is  responsible for the 
Lyman-$\alpha$ forest.

\parn 3. We have investigated the possibility of using a quantitative
comparison of the transverse 
and longitudinal correlation functions (a modified version  
of the Alcock \& Paczy\'nski test) to  constrain  $\Ol$. 
At larges scales ($>$2 arcmin, $>350$ km~s$^{-1}$) the density 
field is in the linear regime. We have thus modelled the peculiar 
velocities using linear theory for the evolution of density fluctuations. 
The linear theory predictions have been fitted 
simultaneously to the transverse and longitudinal  correlations 
of the flux distribution and the procedure has been tested 
on numerical simulations. 
The current data set is not yet large enough
to give significant constraint on  $\Ol$.
However, application of the test to simulated spectra of 
30 pairs at 2, 4.5 (and 7.5) arcmin
recovered  the correct value of $\Ol\pm 0.1$
in 95\% of the realizations.
 If the expected weak correlation at scales larger than 3
arcmin can indeed be detected in larger samples of QSO pairs 
a more sophisticated analysis making more extensive use of 
numerical simulation  should deliver very significant constraints 
on cosmological parameters. Careful calibration with numerical
simulations may also allow to exploit the much stronger correlations 
in the non-linear regime at scales smalller than two arcmin.

\vskip 0.5cm
\noindent{\sl  Acknowledgements:} {This work  was supported  in part  by the
European RTN  program "The Physics of the Intergalactic  Medium" 
\parn P. Petitjean and E.  Rollinde  thank  the Inter  University  Center  
for  Astronomy  and Astrophysics of Pune (IUCAA, India), the National Center for  
Radio Astronomy, TIFR of Pune (NCRA, India), the Astrophysikalisches Intitut Potsdam
(AIP, Germany)  and the Institute of Astronomy of Cambridge (IoA, UK) and
M. Haehnelt thanks the Institut d'Astrophysique de Paris (IAP, France)
for hospitality during the time part of this work was completed.
\parn We thank E.~Thi\'ebaut, 
and  D.~Munro  for  freely  distributing  his  Yorick  programming  language
(available at {\em\tt ftp://ftp-icf.llnl.gov:/pub/Yorick}) which we  used to
implement our algorithm.}
\parn The computational means (NEC-5x5) to do the N-body simulation
were made available to us thanks to the scientific 
council of the Institut du D\'eveloppement et des Ressources 
en Informatique Scientifique (IDRIS).
\parn  We also thank the anonymous referee for useful comments.
\vskip 0.5cm

%

\end{document}